\documentclass[a4paper]{jpconf}

\usepackage{epsfig}

\usepackage{graphicx}

\newcommand{\be}{\begin{equation}}
\newcommand{\ee}{\end{equation}}
\newcommand{\bea}{\begin{eqnarray}}
\newcommand{\eea}{\end{eqnarray}}
\newcommand{\beaa}{\begin{eqnarray}}
\newcommand{\eeaa}{\end{eqnarray}}
\newcommand{\ba}{\begin{array}}
\newcommand{\ea}{\end{array}}
\newcommand{\bit}{\begin{itemize}}
\newcommand{\eit}{\end{itemize}}
\newcommand{\ben}{\begin{enumerate}}
\newcommand{\een}{\end{enumerate}}

\def\lab{\label}

\def\mapbelow#1{\smash{\mathop{\longrightarrow}\limits_{#1}}}

\begin{document}
\title{Dissipation and spontaneous  symmetry breaking in brain dynamics
}

\author{Walter J Freeman${}^{1}$ and Giuseppe Vitiello${}^{2}$}

\address{${}^{1}$ Department of Molecular and Cell Biology\\
University of California, Berkeley CA 94720-3206
USA\\
http://sulcus.berkeley.edu}
\address{${}^{2}$ Dipartimento di Matematica e Informatica, Universit\`a di Salerno,\\
and Istituto Nazionale di Fisica Nucleare, Gruppo Collegato di
Salerno,
I-84100 Salerno, Italy\\
http://www.sa.infn.it/giuseppe.vitiello}



\ead{dfreeman@berkeley.edu,~vitiello@sa.infn.it}

\begin{abstract}
We compare the predictions of the dissipative quantum model of brain
with neurophysiological data collected from electroencephalograms
resulting from high-density arrays fixed on the surfaces of primary
sensory and limbic areas of trained rabbits and cats. Functional
brain imaging in relation to behavior reveals the formation of
coherent domains of synchronized neuronal oscillatory activity and
phase transitions predicted by the dissipative model.
\end{abstract}


\section{Introduction}

In his pioneering work in the first half of the 20th century Lashley
was led to the hypothesis of ``mass action'' in the storage and
retrieval of memories in the brain and observed: ``...Here is the
dilemma. Nerve impulses are transmitted ...from cell to cell through
definite intercellular connections. Yet, all behavior seems to be
determined by masses of excitation...within general fields of
activity, without regard to particular nerve cells... What sort of
nervous organization might be capable of responding to a pattern of
excitation without limited specialized path of conduction? The
problem is almost universal in the activity of the nervous system''
(pp. 302-306  of \cite{Lashley:1948}). Lashley's finding was
confirmed in many subsequent laboratory observations and Pribram
then proposed the analogy between the fields of distributed neural
activity in the brain and the wave patterns in holograms \cite{1}.

Mass action has been confirmed by EEG, by magnetoencephalogram
(MEG), functional magnetic resonance imaging (fMRI), positron
electron tomography (PET), 
and single photon emission computed tomography (SPECT). These
techniques gave observational access to real time imaging of
``patterns of excitation'' and dynamical formation of spatially
extended domains of neuronal fields of activity. The neocortex is
observed to be characterized by the exchangeability of its ports of
sensory input; its ability to adapt rapidly and flexibly to short-
and long-term changes; its reliance on large-scale organization of
patterns of neural activity that mediate its perceptual functions;
the incredibly small amounts of information entering each port in
brief behavioral time frames that support effective and efficient
intentional action and perception
\cite{Bach-y-Rita:1995,Bach-y-Rita}.

None of the following four material agencies which have been
proposed to account for the processes involving large populations of
neurons, appear to be able to explain the observed cortical activity
\cite{FreemanNDN}:

1 - Nonsynaptic transmission is essential for neuromodulation and
diffusion of chemical fields of metabolites providing manifestations
of widespread coordinated firing. It has been proposed
\cite{Bach-y-Rita:1995} as the mechanism for implementation of
volume transmission to answer the question of how  broad and diffuse
chemical gradients might induce phase locking of neural pulse trains
at $ms$ intervals. However, it is too slow to explain the highly
textured patterns and their rapid changes \cite{FreemanNDN}.
Observations \cite{13} show that cortex indeed jumps abruptly from a
receiving state to an active transmitting state. Spatial amplitude
modulated (AM) patterns with carrier frequencies in the beta and
gamma ranges ($12-80 ~Hz$) form during the active state and dissolve
as the cortex returns to its receiving state after transmission.
These state transitions in cortex form frames of AM patterns in few
$ms$, hold them for $80-120 ~ms$, and repeat them at rates in alpha
and theta ranges ($3-12 ~Hz$) of EEG \cite{11}, \cite{13} -
\cite{14}. These patterns appear often to extend over spatial
domains covering much of the hemisphere in rabbits and cats
\cite{12,13}, and over the length of a $64 \times 1$ linear $19 ~cm$
array \cite{7} in human cortex with near zero phase dispersion
\cite{14,15}. Synchronized oscillation of large-scale neuronal
assemblies in beta and gamma ranges have been detected in the
resting state and in motor task-related states of the human brain by
MEG  \cite{Bassett}. The observed high rates of field modulation are
not compatible with mediation of chemical diffusion such as those
estimated in studies of spike timing among multiple pulse trains
(e.g. \cite{Amit:1989,Morrow,Schillen}), of cerebral blood flow
using fMRI  (e.g. \cite{Roland,Varela}), and of spatial patterns of
the distributions of radio-labeled neurotransmitters and
neuromodulators as measured with PET, SPECT and optical techniques.

2 - Electric fields are revealed by the extracellular flow of
dendritic current across the resistance of brain tissue\cite{17}.
Weak extracellular electric currents have been shown to modulate the
firing of neurons  {\it in vitro} and have been postulated as the
agency by which neurons are linked together \cite{Terzuolo}.
However, the current densities required {\it in vivo} to modulate
cortical firing exceed by nearly two orders of magnitude those
currents that are sustained by extracellular dendritic currents
\cite{17,vanHarreveld}.

3 - Magnetic fields of such intensity that they can be measured $4-5
~cm$ above the scalp with MEG are generated by the intracellular
current in palisades of dendritic shafts in cortical columns. The
earth's far stronger magnetic field can be detected by specialized
receptors for navigation in birds and bees \cite{Walker}, leading to
the search for magnetic receptors among cortical neurons (e.g.
\cite{Azanza,Dunn}), so far without positive results.

4 - The combined agency of electric and magnetic fields propagating
as radio waves has also been postulated \cite{Adey}. However,
neuronal radio communication is unlikely, owing to the $80:1$
disparity between electric permittivity and magnetic permeability of
the brain tissue and to the low frequency ($< 100 ~Hz$) and
kilometer wavelengths of electromagnetic radiation at EEG
frequencies.

Thus, neither the chemical diffusion, which is much too slow, nor
the electric field of the extracellular dendritic current nor the
magnetic fields inside the dendritic shafts, which are much too
weak, are the agency of the collective neuronal activity. Lashley's
dilemma remains, thus, still to be explained.

The  dissipative quantum model of brain, which we compare with
laboratory observations in this paper, has been proposed
\cite{Vitiello:1995wv,Vitiello:2001} as an alternative approach to
account for the observed dynamical formation of spatially extended
domains of neuronal synchronized oscillations and of their rapid
sequencing. The  dissipative model explains indeed two main features
of the EEG data \cite{11}: the textured patterns of AM in distinct
frequency bands correlated with categories of conditioned stimuli,
i.e. {\it coexistence} of physically distinct AM patterns, and the
remarkably rapid onset of AM patterns into (irreversible) sequences
that resemble cinematographic frames. Each spatial AM pattern is
described to be consequent to spontaneous breakdown of symmetry
(SBS) triggered by external stimulus and is associated with one of
the emerging unitarily inequivalent ground states. Their sequencing
is associated to the non-unitary time evolution implied by
dissipation, as discussed below. It has to be remarked that the
neuron and the glia cells and other physiological units are {\it
not} quantum objects in the many-body model of brain. This
distinguishes the dissipative quantum model from all other quantum
approaches to brain, mind and behavior. Moreover, the dissipative
model describes the brain, not mental states. Also in this respect
this model differs from those approaches where brain and mind are
treated as if they were a priori identical.

In  Section $2$ and $3$ we briefly summarize the main features of
the original many-body model and its extension to  dissipative
dynamics, respectively. In Section $4$ we comment on the laboratory
observations and their agreement with the dissipative model. We
closely follow Ref. \cite{11} in our presentation. Free energy, the
arrow of time and classicality are discussed in Section $5$ and $6$,
respectively. Conclusions are presented in Section $7$. For the
reader's convenience and for completeness, details of the SBS
mechanism in quantum field theory (QFT) and of the observational
techniques are presented in Appendices A and B, respectively.

\section{The original many-body model}

The dissipative quantum model \cite{Vitiello:1995wv}, on which we
focus our attention in this paper, extends the original quantum
model of brain to the dissipative dynamics intrinsic to the brain
functional activity. The quantum model of brain, here briefly
summarized, was proposed in 1967 by Ricciardi and Umezawa \cite{UR}
and further developed by Stuart, Takahashi and Umezawa
\cite{Stuart}, see also \cite{CH}. It was formulated in order to
provide a solution to Lashley's  dilemma. The model is primarily
aimed to the description of memory storing and recalling. Umezawa
explains the motivation for using the QFT formalism of many-body
physics \cite{UC}: "In any material in condensed matter physics any
particular information is carried by certain ordered patterns
maintained by certain long range correlations mediated by massless
quanta. It looked to me that this is the only way to memorize some
information; memory is a printed pattern of order supported by long
range correlations..."

The main ingredient of the model is thus the mechanism of SBS by
which long range correlations (the Nambu-Goldstone, briefly NG,
boson modes) are dynamically generated (see the Appendix A). Water
constitutes more than  $80\% $ to brain mass, and in the many-body
model it is therefore expected to be a major facilitator or
constraint on brain dynamics. The symmetry which gets broken is the
rotational symmetry of the electric dipole vibrational field of the
water molecules and of other biomolecules present in the brain
structures \cite{Vitiello:1995wv,DelGiudice:1985,Jibu}.  The quantum
variables are identified with those of the electric dipole
vibrational field and with the associated NG modes, named the dipole
wave quanta (DWQ). These are dynamically created and do not derive
from Coulomb interaction.

If the cortex is at or near a singularity (see Section 3), the
external input or stimulus acts on the brain as a trigger for the
breakdown of the dipole rotational symmetry. As a consequence long
range correlation is established by the coherent condensation of DWQ
bosons. SBS guarantees the change of scale, from the microscopic
dynamics to the macroscopic order parameter field. The density value
of the condensation of DWQ in the ground state (also called vacuum
state) acts as a {\it label} classifying the state and thus the
memory thereby created. The stored memory  is not a representation
of the stimulus, nor is it a collection of stimulus features.
Indeed, a specific feature of the SBS mechanism in QFT is that the
ordered pattern generated is controlled by the inner dynamics of the
system, not by the external field (stimulus) whose only effect is
the breakdown of the symmetry. This aspect of the model perfectly
agrees with laboratory observations (see Sections 3 and 4).

The recall of the recorded information occurs under the input of a
stimulus capable of exciting DWQ out of the corresponding ground
state. In the model,  such a stimulus is called ``similar" to the
one responsible for the memory recording \cite{Stuart}. Similarity
is not an intrinsic property of the stimuli. Rather, it refers to
their effects on the brain, namely inducing the formation or
excitation of ``similar" ordered pattern(s).

One shortcoming of the many-body model in its original form is that
any subsequent stimulus would cancel the previously recorded memory
by renewing the SBS process and the consequent DWQ condensation,
thus printing the new memory over the previous one (``memory
capacity problem"). Moreover, the model fails in explaining the
observed coexistence of AM patterns and their irreversible time
evolution. These problems are solved by endorsing the original
many-body model with dissipative dynamics
\cite{Vitiello:1995wv,Vitiello:2001}, accounting for the fact that
the brain is an open system in permanent interaction with its
environment.

\section{The dissipative many-body model}

\vspace{.8mm}

\subsection{Coherent states}

The details of the coupling of the brain with environment are very
intricate and variable, and thus they are difficult to be
characterized and measured. The external stimulus on the brain {\it
selects} one vacuum state among infinitely many of them, unitarily
inequivalent with each other (see Appendix A). The selection of the
vacuum is  what happens in the process of SBS. The selected vacuum
carries the {\it signature} (memory) of the reciprocal
brain--environment influence at a given time under given boundary
conditions. A change in the brain--environment interaction changes
the choice of the vacuum: the brain evolution through the vacuum
states thus reflects the evolution of the coupling of the brain with
the surrounding world. The condensate of DWQ in the vacuum is
assumed to be the quantum substrate of the observed AM patterns. In
agreement with observations, the dissipative dynamics allows
(quasi-)non-interfering degenerate vacua with different condensates.
This corresponds to different AM patterns
and (phase) transitions among them (AM pattern sequencing). These
features could not be described in the framework of the original
many-body model. By exploiting the existence of infinitely many
inequivalent modes in QFT, the dissipative model allows a huge
memory capacity. This can be seen as follows.

In QFT the canonical quantization of a dissipative system requires
that the environment in which the system is embedded must  also be
included in the formalism. This is achieved by describing the
environment as the time-reversed image of the system, and this is
realized by doubling  the  system's degrees of freedom
\cite{Celeghini:1992yv}. In the dissipative quantum model, the brain
dynamics is indeed described in terms of an infinite collection of
damped harmonic oscillators $a_{\kappa}$ (a simple prototype of a
dissipative system) representing the boson DWQ modes
\cite{Vitiello:1995wv} and by the ${\tilde a}_{\kappa}$ modes which
are the time-reversed mirror image  of the $a_{\kappa}$ modes. The
doubled modes ${\tilde a}_{\kappa}$ represent the environment. The
role of the ${\tilde a}_{\kappa}$ system is to restore energy
conservation by balancing the (in-/out-)energy fluxes. The label
$\kappa$ generically denotes degrees of freedom such as, e.g.,
spatial momentum, etc.
\cite{Vitiello:1995wv,Celeghini:1992yv,Alfinito:2000ck}.

The $a_{\kappa}$ and ${\tilde a}_{\kappa}$ modes are massless NG
modes. The system Hamiltonian is invariant under the dipole
rotations (described by the $SU(2)$ group). The breakdown of this
rotational symmetry is induced by the external stimulus and this
leads to the dynamical generation of DWQ $a_{\kappa}$. Their
condensation in the ground state is then constrained by inclusion of
the mirror modes ${\tilde a}_{\kappa}$ in order to account for the
system dissipation. The system ground state is indeed  not invariant
under (continuous) time translation symmetry. As a result we have
energy non-conservation and irreversible time evolution  for the
$a_{\kappa}$ system. One can show
\cite{Celeghini:1992yv,Alfinito:2000ck} that the external stimulus
formally represents the coupling strength between the $a_{\kappa}$
and the ${\tilde a}_{\kappa}$ modes.

Although the living brain operates far from equilibrium, it evolves
in time through a sequence of states where the energy fluxes and
heat exchanges at the system-environment interface are balanced:
$E_{syst} - E_{env} \equiv E_{0} = 0$. This energy balance is
manifested in the regulation of mammalian brain temperature. The
balanced nonequilibrium system state, denoted by ${|0\rangle}_{\cal
N}$, is thus the system vacuum or ground state. At some arbitrary
initial time $t_{0} = 0$, the Hamiltonian prescribes
\cite{Vitiello:1995wv} that $E_{0} = \sum_{\kappa} \hbar
\Omega_{\kappa} ({\cal N}_{a_{\kappa}} - {\cal N}_{{\tilde
a}_{\kappa}})= 0$, where $\Omega_{\kappa}$ is the common frequency
of $a_{\kappa}$ and ${\tilde a}_{\kappa}$ modes.  This implies that
the "memory state" ${|0\rangle }_{\cal N}$ is a condensate of an
{\it equal number} of modes $a_{\kappa}$ and mirror modes ${\tilde
a}_{\kappa}$ for any $\kappa$: ${\cal N}_{a_{\kappa}} - {\cal
N}_{{\tilde a}_{\kappa}} = 0$ \footnote{Let   $\{ | {\cal
N}_{a_{\kappa}} , {\cal N}_{\tilde a_{\kappa}} \rangle \}$ be the
set of simultaneous eigenvectors of ${\hat N}_{a_{\kappa}} \equiv
a^{\dagger}_{\kappa} a_{\kappa}$ and ${\hat N}_{\tilde a_{\kappa}}
\equiv {\tilde a_{\kappa}}^{\dagger} {\tilde a_{\kappa}}$, with
${\cal N}_{a_{\kappa}}$ and ${\cal N}_{\tilde a_{\kappa}}$
non-negative integers. Then $|0\rangle_{0} \equiv | {\cal
N}_{a_{\kappa}} = 0 , {\cal N}_{\tilde a_{\kappa}} = 0 \rangle$
denotes the state annihilated by $a_{\kappa}$ and by ${\tilde
a_{\kappa}}$: $a_{\kappa} |0\rangle_{0} = 0 = {\tilde
a_{\kappa}}|0\rangle_{0} $ for any $\kappa$.}. We have ${}_{\cal
N}\langle 0 | 0\rangle_{\cal N} = 1 ~ \forall ~ \, {\cal N}$, where
${\cal N}$ denotes the set of integers defining the "initial value"
of the condensate, ${\cal N} \equiv \{ {\cal N}_{a_{\kappa}} = {\cal
N}_{{\tilde a}_{\kappa}}, \forall ~ \kappa, at~~  t_{0} = 0 \}$, as
the {\it order parameter} associated with the information recorded
at time $t_{0} = 0$.

Clearly, balancing $E_{0}$ to be zero does not fix the value of
either $E_{a_{\kappa}}$ or $E_{{\tilde a}_{\kappa}}$ for any
$\kappa$. It only fixes, for any $\kappa$, their difference.
Therefore, at $t_{0}$ we may have infinitely many perceptual states,
each of which is in one-to-one correspondence to a given  set ${\cal
N}$. The dynamics ensures that the number $ ( {\cal N}_{a_{\kappa}}
- {\cal N}_{{\tilde a}_{\kappa}})$ is a constant of motion for any
$\kappa$ (see \cite{Vitiello:1995wv}). The average number ${\cal
N}_{a_{\kappa}}$ is given by
\be\lab{num} {\cal N}_{a_{\kappa}} =
 {_{\cal N}}\langle  0| a_{\kappa}^{\dagger} a_{\kappa}{|0\rangle}
 _{\cal N} = \sinh^{2} \theta_{\kappa}~,
\ee
where ${\theta}_{\kappa}$ is a transformation parameter. The
$\theta$-set, $\theta \equiv \{ {\theta}_{\kappa}, \forall ~ \kappa,
at~~  t_{0} = 0 \}$, is related to the $\cal N$-set, ${\cal N}
\equiv \{ {\cal N}_{a_{\kappa}} = {\cal N}_{{\tilde a}_{\kappa}},
\forall ~ \kappa, at~~ t_{0}=0 \}$, by Eq. (\ref{num}). We also use
the notation ${\cal N}_{a_{\kappa}}(\theta) \equiv {\cal
N}_{a_{\kappa}}$ and ${|0(\theta) \rangle} \equiv {|0 \rangle}_{\cal
N}$. 
The $\theta$-set is conditioned by the requirement that $a_{\kappa}$
and ${\tilde a}_{\kappa}$ modes satisfy the Bose-Einstein
distribution:
\be\lab{bose} {\cal N}_{a_{\kappa}}(\theta) = \sinh^{2}
\theta_{\kappa}
 = {1\over{{\rm e}^{\beta
 E_{\kappa}} - 1}} \quad ,
\ee
where  $ {\beta {\equiv} {1\over{k_{B} T}}}$  is the inverse
temperature at time $t_{0}=0$ ($k_{B}$ is Boltzmann's constant).
${|0\rangle}_{\cal N}$ is thus recognized to be a finite temperature
state and it can be shown to be a squeezed coherent state
\cite{Vitiello:1995wv,Umezawa:1993yq,perelomov,TakUmez}.

The spaces $\{{|0\rangle}_{\cal N} \}$ and $\{{|0 \rangle}_{\cal
N'}\}$ are  unitarily inequivalent with each other for different
labels ${\cal N} \neq {\cal N'}$ in the infinite volume limit. This
is expressed by the relation:
\be\lab{ort1} {}_{\cal N}\langle 0 | 0 \rangle_{\cal N'} \mapbelow{V
\rightarrow \infty} 0  \quad \quad \forall ~ \, {\cal N},~ {\cal N'}
~,~\quad {\cal N} \neq {\cal N'}~. \ee
We have therefore infinitely many unitarily inequivalent spaces of
states $\{{|0\rangle}_{\cal N} \}$. The set of all these spaces
constitutes the whole space of states. A huge number of sequentially
recorded memories may thus {\it coexist} without destructive
interference since infinitely many vacua ${|0\rangle}_{\cal N}$ are
independently accessible. In contrast to the non-dissipative model,
recording the memory $\cal N'$ does not necessarily produce
destruction of a previously printed memory ${\cal N} \neq {\cal
N'}$; this is the meaning of the non-overlapping modes in the
infinite volume limit expressed by Eq. (\ref{ort1}). Through the
doubled degrees of freedom ${\tilde a}_{\kappa}$, dissipation allows
the possibility of a huge memory capacity by introducing the $\cal
N$-labeled ``replicas" of the ground state. The dissipative model
thus predicts the existence of textures of AM patterns (cf. Sec. 4),

These patterns are represented by order parameters that are stable
against quantum fluctuations. This is a manifestation of the {\it
coherence} of the DWQ boson condensation. In this sense,  the order
parameter is a macroscopic observable and the state
${|0\rangle}_{\cal N}$ provides an example of macroscopic quantum
state. The change of scale (from microscopic to macroscopic) is
dynamically achieved through the SBS leading to boson condensation.

\vspace{.4mm}

\subsection{Phase transitions}

The brain (ground) state may be represented as the collection (or
the superposition) of the full set of states ${|0\rangle}_{\cal N}$,
for all $\cal N$. In the memory space or the {\it brain state
space},  each representation $\{{|0\rangle}_{\cal N}\}$ denotes a
physical phase of the system and may be conceived as a ``point''
identified by a specific $\cal N$-set (or $\theta$-set). In the
infinite volume limit, points corresponding to different $\cal N$
(or $\theta$) sets are distinct points (do not overlap, cf. Eq.
(\ref{ort1})).  The brain in relation to the environment may occupy
any of the ground states, depending on how the $E_{0} = 0$ balance
is approached. Or, it may be in any state that is a collection or
superposition of these brain-environment equilibrium ground states.
Under the influence of one or more stimuli (acting as  control
parameters), the system may shift from ground state to ground state
in this collection (from phase to phase), namely it may undergo an
extremely rich sequence of phase transitions, leading to the
actualization of a sequence of dissipative structures formed by AM
patterns (see Sec. 4).

Let $|0(t)\rangle_{\cal N} $ denote the state $| 0\rangle_{\cal N}$
at time $t$ specified by the initial value ${\cal N}$,  at $t_{0} =
0$. We have $ {}_{\cal N}\langle0(t) | 0(t)\rangle_{\cal N} = 1,
\quad \forall ~t$. We can show that
\be \lim_{t\to \infty} {}_{\cal N}\langle0(t) | 0\rangle_{\cal N} \,
\propto \lim_{t\to \infty}
 \exp{\left ( -t  \sum_{\kappa}  \Gamma_{\kappa}  \right )} = 0 ~,
 \label{(12)}
\ee
provided $ {\sum_{\kappa} \Gamma_{\kappa} > 0}$. In the infinite
volume limit we have (for $ {\int \! d^{3} \kappa \,
\Gamma_{\kappa}}$ finite and positive)
\be
{}_{\cal N}\langle0(t) | 0(t') \rangle_{\cal N}  \mapbelow{V
\rightarrow \infty} 0 \quad \quad \forall \, t\, , t' ~ , \quad t
\neq t' ~.
 \label{(13)}
\ee

The time evolution of the  state $|0(t)\rangle_{\cal N}$ is thus
represented as the trajectory starting with ``initial condition"
specified by the $\cal N$-set in the space $\{ |0(t)\rangle_{\cal N}
\}$. In a pictorial way we could say that the state
$|0(t)\rangle_{\cal N}$ provides the ``instantaneous picture" of the
system at each instant of time $t$, or the ``photograph" at $t$ in a
cinematographic sequence.

Time--dependence of the DWQ frequency implies that higher momentum
$\kappa$-components of the $\cal N$-set possess longer life--times.
Momentum is proportional to the inverse distance over which the mode
propagates, thus modes with a shorter range of propagation (more
``localized" modes) survive longer. On the contrary, modes with a
longer range of propagation decay sooner.

As a result, condensation domains of different finite sizes with
different degrees of stability are predicted by the model
\cite{Alfinito:2000ck}. They are described by the condensation
function $f(x)$ which acts as a ``form factor" specific for the
considered domain
\cite{Umezawa:1993yq,Alfinito:2002a,Alfinito:2002b}. $f(x)$ has to
carry some topological singularity in order for the condensation
process to be physically detectable. A regular function $f(x)$ would
produce a condensation which could be easily ``washed" out
(``gauged" away by a convenient gauge transformation). In a similar
way, the phase transition from one space to another (inequivalent)
space can only be  induced by  a singular condensation function
$f(x)$. This  explains why topologically non trivial extended
objects, such as vortices, appear in phase transitions
\cite{Umezawa:1993yq,Alfinito:2002a,Alfinito:2002b}. Phase
transitions driven by boson condensation are always associated with
some singularity (indeterminacy) in the field phase at the phase
transition point \cite{11X}. This model feature accounts for a
crucial mechanism observed in laboratory experiments: the event that
initiates a perceptual phase transition is an abrupt decrease in the
analytic power of the background activity to near zero.

\section{Observation in cortical dynamics}

The high spatial resolution required to measure AM pattern textures
in brain activity is achieved by using high-density electrode
arrays, fixed on the scalp or the epidural surface of cortical
areas, and fast Fourier transform (FFT) \cite{7,FreemanSilbergeld}.
The set of $n$ amplitudes squared from an array of $n$ electrodes
(typically $64$) defines a feature vector, ${\bf A^{2}}(t)$, of the
spatial pattern of power at time $t$. The vector specifies a point
on a dynamic trajectory in {\it brain state space}, conceived as the
collection of all possible (essentially infinitely many) brain
states. The measurement of $n$ EEG signals defines a finite
$n$-dimensional subspace, so the point specified by ${\bf A^{2}}(t)$
is unique for a spatial AM pattern of an aperiodic carrier wave.
Similar AM patterns form a cluster in $n$-space, and multiple
patterns form either multiple clusters or trajectories with large
Euclidean distances between the digitizing steps in $n$-space. A
cluster with a verified behavioral correlate denotes an {\it ordered
AM pattern}: when the trajectory of a sequence of points enters into
a cluster, that location in state space signifies increased order
from the perspective of an intentional state of the brain, owing to
the correlation with a conditioned stimulus (for further details see
Appendix B).

\begin{figure}
\begin{center}
\includegraphics [width=20pc] {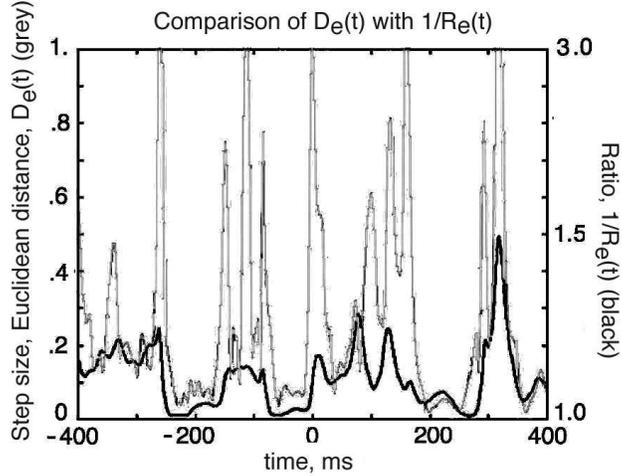}
\end{center}
\caption{\nonumber \small \noindent  
The sharp spikes (gray, De(t)) show the rate of change in spatial AM
pattern. The lower curve (black, the inverse of Re(t), a measure of
synchrony) shows that re-synchronization precedes the emergence of
spatial order and also the increase in power in each frame (see also
Fig. B1).}
\end{figure}

The inverse of the absolute value of the step size between
successive values of $D_{e}(t) = |{\bf A^{2}}(t) - {\bf
A}^{2}(t-1)|$ provides a scalar index of the order parameter.
Indeed, small steps in Euclidean distances, $D_{e}(t)$ (higher
spikes in Fig. 1) indicate pattern amplitude stability. Pattern
phase stability can be characterized by calculating the ratio,
$R_{e}(t)$, of the temporal standard deviation of the mean filtered
EEG to the mean temporal standard deviation of the $n$ EEGs
\cite{9,10} (lower curve in Fig. 1). $R_{e}(t) = 1$ when the
oscillations are complitely synchronized. When $n$ EEGs are totally
desynchronized, $R_{e}(t)$ approaches one over the square root of
the number of digitizing steps in the moving time window. It was
experimentally found that  $R_{e}(t)$ rises rapidly within a few
$ms$ after a phase discontinuity and several $ms$ before the onset
of a marked increase in mean analytic amplitude, $\underline{A}(t)$.

The succession of the high and low values of $R_{e}(t)$ reveals the
episodic emergence and dissolution of synchrony; since cortical
transmission of spatial patterns is most energy-efficient when the
dendritic currents are most synchronized, $R_{e}(t)$ can be adopted
as an index of cortical {\it efficiency} \cite{Haken:1996}.
Re-synchronized oscillations in the beta range near zero lag
commonly recur at rates in the theta range. They cover substantial
portions of the left cerebral hemisphere \cite{8} in some instances
appearing to exceed the length of the recording array (19 cm) on the
scalp above the human brain.

Assuming that the phenomenological order parameter ${\bf A^{2}}(t)$
corresponds to the order parameter ${\cal N}$ introduced in the
dissipative model, the trajectories described by the time dependent
vector ${\bf A^{2}}(t)$ in the brain state space have their quantum
image in the time evolution in the spaces $\{|0\rangle_{\cal N}\}$.

Considering the common frequency $\Omega_{\kappa}(t)$ for the
$a_{\kappa}$ and ${\tilde a}_{\kappa}$ modes (cf. Eq. (8) in
\cite{Alfinito:2000ck}) in the dissipative model, the duration, size
and power of AM patterns are predicted to be decreasing functions of
the carrier wave number $\kappa$. This is confirmed by the
observations. Carrier waves in the gamma range ($30-80 ~Hz$) show
durations seldom exceeding $100 ~ms$, diameters seldom exceeding $15
~mm$; and low power in a $1/f^{a}$ power law as a function of
frequency. Carrier frequencies in the beta range ($12-30 ~Hz$) show
durations often exceeding $100 ~ms$; estimated diameters large
enough to include multiple primary sensory areas and the limbic
system; and they have
greater power. 

The reduction in the amplitude of the spontaneous background
activity induces a brief state of instability, depicted as a null
spike \cite{Freeman:2007a}, in which the significant pass band of
the ECoG is near to zero and its phase is undefined, as indeed
predicted by the dissipative model. The cortex can be driven across
the phase transition process to a new AM pattern by the stimulus
arriving at or just before this state. When considering the
normalized amplitude defined as the AM pattern divided by the mean
amplitude, which is input dependent, one observes that, again in
agreement with prediction of the dissipative model, such a
normalized response amplitude depends not on the input amplitude,
but on the intrinsic state of the cortex, specifically the degree of
reduction in the power and order of the background brown noise. The
null spike in the band pass filtered brown noise activity is
conceived as a {\it shutter} that blanks the intrinsic background.
At very low analytic amplitude when the analytic phase is undefined,
the system, under the incoming weak sensory input, may re-set the
background activity in a new AM frame, if any, formed by
reorganizing the existing activity, not by the driving of the
cortical activity by input (except for the small energy provided by
the stimulus that is required to force the phase transition (and
select an attractor, see below)). The decrease ({\it shutter})
repeats aperiodically in the theta or alpha range, independently of
the repetitive sampling of the environment by limbic input and
allows opportunities for phase transitions.

In conclusion, the reduction in activity constitutes a singularity
in the dynamics at which the phase is undefined, in agreement with
the dissipative model requiring the singularity of the boson
condensation function.  The power is not provided by the input,
exactly as the dissipative model predicts, but by the pyramidal
cells, which explains the lack of invariance of AM patterns with
invariant stimuli \cite{Freeman:2007a}.

Finally, we note that another possible way to break the symmetry in
QFT is to modify the dynamical equations by adding one or more terms
that are explicitly not consistent with the symmetry transformations
(i.e., are not symmetric terms). This is called {\it explicit}
breakdown of symmetry. The system is forced by the external action
into a specific non-symmetric state that is determined by the
imposed breaking term. The explicit breakdown  fits well with {\it
event-related potentials} (ERP) observed as the response of the
cortex to perturbations, such as an electric shock, sensory click,
flash, or touch. By resorting to stimulus-locked averaging across
multiple presentations in order to remove or attenuate the
background activity, the location, intensity and detailed
configuration of the ERP are  predominantly determined by the
stimulus; so  ERP signals can be used as evidence for processing by
the cortex of exogenous information. In contrast, in SBS pattern
configurations are determined from information that is endogenous
from the memory store.

\section{The thermal connection: free energy and the arrow of time}

In Section $3$ we have seen that the brain states ${|0\rangle}_{\cal
N}$ are finite temperature states. This shows the intrinsically
thermal nature of brain dynamics which we analyze  further in the
present Section.

In the dissipative model the free energy functional for the
$a_{\kappa}$ modes is given by \cite{Vitiello:1995wv}
\be {\cal F}_{a} \equiv {}_{\cal N}\langle0(t)| \Bigl ( H_{a}
-{1\over{\beta}} S_{a} \Bigr ) |0(t)\rangle_{\cal N}  \quad ,
\lab{(20)} \ee
with time-dependent inverse temperature $ {\beta (t) = {1\over{k_{B}
T(t)}}}$. $S_{a}$ is the entropy operator  given by
\be\lab{MMt} S_{a} \equiv - \sum_{\kappa} \Bigl \{
a_{\kappa}^{\dagger} a_{\kappa} \ln \sinh^{2}  {\Theta}_{\kappa} -
a_{\kappa} a_{\kappa}^{\dagger} \ln \cosh^{2} {\Theta}_{\kappa}
\Bigr \}~, \ee
where $\Theta_{\kappa} \equiv \Gamma_{k} t - \theta_{k}$. Here
$\Gamma_{\kappa}$ is the damping constant and $\theta_{\kappa}$ is
the transformation parameter introduced in Eq. (\ref{num}).
$S_{\tilde a}$ is obtained by replacing $a_{\kappa}$ and
$a_{\kappa}^{\dagger}$ with ${\tilde a}_{\kappa}$ and ${\tilde
a}_{\kappa}^{\dagger}$, respectively, in (\ref{MMt}). $H_{a}$
denotes the Hamiltonian at $t = t_{0}$ relative to the
$a_{\kappa}$-modes only, ${H_{a} = \sum_{k} E_{k} a_{k}^{\dagger}
a_{k}}$, with $E_{k} \equiv \hbar \Omega_{k}(t_{0})$. For the
complete system $a-{\tilde a}$, the difference ~$(S_{a} - S_{\tilde
a})$~ is constant in time: $ [\, S_{a} - S_{\tilde a} , {\cal
H}^{\prime} ] = 0$. The stationarity condition to be satisfied at
each time $t$ by the state $|0(t)\rangle_{\cal N} $ is ~${{\partial
{\cal F}_{a}}\over{\partial \Theta_{k}}} = 0 , ~ \forall ~ k $,
which, for $\beta (t)$ slowly varying in time, i.e. $ {{{\partial
\beta}\over{\partial t}} = - {1\over{k_{\tilde a} T^{2}}} {{\partial
T}\over{\partial t}} \approx ~0}$, gives the Bose-Einstein
distribution
\be {\cal N}_{a_{k}}(\theta,t) =  {1\over{{\rm e}^{\beta (t) E_{k}}
- 1}} \quad . \lab{(22)} \ee

The changes in the energy $ {E_{a} \equiv \sum_{k} E_{k} {\cal
N}_{a_{k}}}$  and in the entropy  ${\cal S}_{a}(t) = \langle0(t)|
S_{a} |0(t)\rangle_{\cal N}$ are given by
\be d E_{a} = \sum_{k} E_{k} \dot{\cal N}_{a_{k}} d t =
 {1\over{\beta}} d {\cal S}_{a}  \quad .
  \lab{(23)}
\ee
Provided that changes in inverse temperature are slow, the
minimization of the free energy thus holds at any $t$:
\be  d {\cal F}_{a} = d E_{a} - {1\over{\beta}} d {\cal S}_{a} = 0
\quad . \lab{(24)} \ee
The time-evolution of the state $|0(t)\rangle_{\cal N}$  at finite
volume $V$ can be shown \cite{Vitiello:1995wv,Celeghini:1992yv} to
be controlled by the entropy variations, which reflects the
irreversibility of time-evolution (breakdown of time-reversal
symmetry) characteristic of dissipative systems. This corresponds to
the choice of a privileged direction in time-evolution called {\it
arrow of time}.

Eq. (\ref{(23)}) shows that the change in time of the condensate,
i.e. of the order parameter, turns  into heat dissipation
$dQ={1\over{\beta}} dS_{a}$. Therefore the ratio of the rate of free
energy dissipation to the rate of change of the order parameter is a
good measure of the ordering stability. In terms of laboratory
observations the rate of change of the order parameter is specified
by the Euclidean distance $D_{e} (t)$ between successive points in
the $n$-space. $D_{e}(t)$ takes large steps between clusters,
decreases to a low value when the trajectory enters a cluster, and
remains low for tens of $ms$ within a frame (Fig. 1). Therefore
$D_{e}(t)$ serves as a measure of the spatial AM pattern stability.

It was found \cite{12,13} that the best predictor for the onset of
ordered AM patterns is the {\it pragmatic information} index
$H_{e}(t)$, so named after Atmanspacher and Scheingraber
\cite{Atmanspacher:1990}, given by the ratio of the rate of free
energy dissipation $\underline{A}^{2}(t)$ to the rate of change of
the order parameter represented by $D_{e}(t)$ (because $D_{e}(t)$
falls and $\underline{A}^{2}(t)$ rises with wave packet evolution):
$$
H_{e}(t) = \frac{\underline{A}^{2}(t)}{D_{e}(t)} ~.
$$
Measurements showed that typically the rate of change in the
instantaneous frequency $\omega (t)$ was low in frames that
coincided with low $D_{e}(t)$ indicating stabilization of frequency
as well as AM pattern. Between frames $\omega (t)$ often  increased
several times or decreased even below zero in interframe breaks that
repeat at rates in the theta or alpha range of the EEG \cite{8}
({\it phase slip} \cite{Pikovsky}).

We observe that the mirror ${\tilde a}_{\kappa}$ modes account
\cite{Srivastava:1995yf,Pessa:2003} for Brownian quantum noise due
to the fluctuating random force in the system-environment coupling.
Such a noise is responsible for the fact that the state
${|0\rangle}_{\cal N}$ is an entangled state \cite{11}, the entropy
operator providing a measure of the entanglement ($a_{\kappa}$ and
${\tilde a}_{\kappa}$ modes are entangled modes). In other words,
the brain processes are inextricably dependent on the quantum noise
in the fluctuating random force in the brain-environment coupling.
There is a permanent brain-environment entanglement. This feature
seems to model the observed continual perturbations involving all
areas of neocortex by other parts of the brain, including inputs
from the sensory receptors that are relayed mainly through the
thalamus, and the catastrophic disruptions of brain function that
result from prolonged sensory deprivation. These continuous
perturbations give rise to myriads of local phase transitions, which
are quenched as rapidly as they are formed, thereby maintaining the
entire cortex in a robust state of conditional stability
(metastability \cite{Kelso,Bressler,Fingelkurts:2004}).  An
interesting question is whether such a regime might conform to
self-organized criticality \cite{7,13,53,81,82} (the mean firing
rate of neurons, homeostatically maintained by mutual excitation
everywhere by thresholds and refractory periods, would play the
r\^ole of the critical variable corresponding to angle in
self-organized criticality \cite{10}). It is indeed interesting
that, in a model \cite{deArcangelis:2006a} based on self-organized
criticality combined with synaptic plasticity in a neural network,
the average power spectrum computed as a function of frequency
exhibits a power law behavior with the same exponent as found in
medical EEG power spectra \cite{FreemanSilbergeld, Novikov:1997a}.

\section{Classicality and attractor landscapes: the classical blanket}

One of the merits of the dissipative many-body model is the
possibility
\cite{Vitiello:1995wv,Vitiello:2001,Alfinito:2000ck,Pessa:2003} of
deriving from the microscopic dynamics the classicality of the
trajectories representing the time-evolution of the state
$|0(t)\rangle_{\cal N}$ in the brain state space. These trajectories
are found to be deterministic chaotic trajectories
\cite{Pessa:2003,Vitiello:2003me}. This is a particularly welcome
feature of the model since observed  changes in the order parameter
become susceptible to be described in terms of trajectories on
attractor landscapes. One can show these trajectories  are {\it
classical}  and that

$i)$~ they are bounded and  do not intersect themselves
(trajectories are not periodic).

$ii)$~~there are no intersections between trajectories specified by
different initial conditions.

$iii)$ trajectories of different initial conditions  diverge.

Although  property $ii)$ implies that no {\it confusion} or
interference arises among different memories, even as time evolves,
states with different $\cal N$ labels may have non--zero overlap
(non-vanishing inner products) in realistic situations of finite
volume. This means that some {\it association} of memories becomes
possible: at a ``crossing" point between two, or more than two,
trajectories, one can ``switch" from one of them  to another one.
This reminds us of the ``mental switch" occurring during particular
perceptual and motor tasks \cite{Kelso,C} as well as during free
associations in memory tasks \cite{D}.

One can derive \cite{Pessa:2003} from  property $iii)$  that the
difference between $\kappa$--components of the sets $\cal N$ and
$\cal N'$ may become zero at a given time $t_{\kappa}$. However, the
difference between the sets $\cal N$ and $\cal N'$ does not
necessarily become zero. The $\cal N$-sets are made up of a large
number (infinite in the continuum limit) of ${\cal
N}_{a_{\kappa}}(\theta,t)$ components, and they are different even
if a finite number (of zero measure) of their components are equal.
On the contrary, for very small ${\delta \theta_{\kappa}}$, suppose
that $\Delta t \equiv \tau_{max} -\tau_{min}$, with $\tau_{min}$ and
$\tau_{max}$ the minimum and the maximum, respectively, of
$t_{\kappa}$, {\it for all} $\kappa$'s, be ``very small''. Then the
$\cal N$-sets are ``recognized'' to be ``almost'' equal in such a
$\Delta t$.  Thus we see how in the ``recognition'' (or recall)
process it is possible that ``slightly different'' ${\cal
N}_{a_{\kappa}}$--patterns  are ``identified'' (recognized to be the
``same pattern'' even if corresponding to slightly different
inputs). Roughly, $\Delta t$ provides a measure of the ``recognition
time''.

The deterministic chaotic motion described by  $i)$--$iii)$ takes
place in the space of the parameters labeling the system ground
state. It is low dimensional and noise-free. In a more realistic
framework, the motion must be conceived as high-dimensional, noisy,
engaged and time-varying.  Nevertheless, it is remarkable that, at
the present stage of our research, the dissipative model predicts
that the system trajectories through its physical phases may be
chaotic \cite{Pessa:2003} and itinerant through a chain of
``attractor ruins" \cite{Tsuda}, embedded in a set of attractor
landscapes \cite{Skarda} accessed serially or merely approached in
the coordinated dynamics of a metastable state
\cite{Srivastava:1995yf,Bressler,Bressler:2001a,Fingelkurts:2004,Fingelkurts:2001}.
The manifold on which the attractor landscapes sit covers as a {\it
classical blanket} the quantum dynamics going on in each of the
representations of the CCR's (the AM patterns recurring at rates in
the theta range ($3-8 ~Hz$)).

We propose conditioned stimuli {\it select} a basin of attraction in
the primary sensory cortex to which it converges ({\it abstraction}
by deletion of nonessential information), often with very little
information as in weak scents, faint clicks, weak flashes. The
astonishingly low requirements for information in high-level
perception have been amply demonstrated by recent accomplishments in
sensory substitution \cite{Cohen,VonMelchner, Bach-y-Rita}. There is
an indefinite number of such basins  forming a pliable and adaptive
attractor landscape in each sensory cortical area. Each attractor
can be selected by a stimulus that is an instance of the category
({\it generalization}) that the attractor implements by its AM
pattern. The waking state consists of a collection of potential
states, any one of which (but only one at a time) can be realized
through a phase transition. The variety of these highly textured,
latent AM patterns, their exceedingly large diameters in comparison
to the small sizes of the component neurons and the extraordinarily
rapid temporal sequence in the neocortical phase transitions by
which they are selected, provide the principal justification for
exploring the interpretation of nonlinear brain dynamics in terms of
many-body theory and multiple ground states.

\section{Concluding remarks}

Our discussion in this paper leads us to conclude that the
dissipative quantum model of brain predicts two main features
observed in  neurophysiological data: the coexistence of physically
distinct AM patterns correlated with categories of conditioned
stimuli and the remarkably rapid onset of AM patterns into
irreversible sequences that resemble cinematographic frames. Each
spatial AM pattern is described to be consequent to the spontaneous
breakdown of symmetry triggered by an external stimulus and is
associated with one of the unitarily inequivalent ground states of
QFT. Their sequencing is associated to the non-unitary time
evolution implied by dissipation. There are many open questions
which remain to be answered. For example, the analysis of the
interaction between the boson condensate and the details of
electrochemical neural activity, or the problems of extending the
dissipative many-body model to account for higher cognitive
functions of the brain need much further work.

One peculiar property of quantum field dynamics, which makes it so
successful in the description of many-body systems with different
phases, and which motivates us to apply it to brain dynamics, is
that there are many stability  ranges, each one characterizing a
specific phase of the system with specific physical properties that
differ from phase to phase (in the brain: from each observed AM
pattern to the next). If the dynamical regime is characterized by a
range of parameter values which does not allow SBS, the system does
not perceptibly or meaningfully react (as in sleep to weak stimuli).
When one or more control parameters, such as the strength of action
at one class of synapses in the cortical pool under the influence of
the weak external stimulus, or even by indeterminate drift, exceeds
the range of stability where the system originally sits, a
transition is induced to another stability parameter range. It
differs from the previous one in that it now allows SBS and the
appearance of order (as in arousal from deep sleep). Contrarywise,
the loss of order as in shutting down under anesthesia or in deep
sleep corresponds to symmetry recovery or restoration, the
formlessness of background activity or in the extreme the loss of
activity in the case of brain death.

The concept of the DWQ boson carrier discussed above enables an
orderly and inclusive description of the phase transition that
includes all levels of the microscopic, mesoscopic, and macroscopic
organization of  cerebral patterns.  The hierarchical structure
extending from atoms to the whole brain and outwardly into
engagement of the subject with its environment in the
action-perception cycle is the essential basis for the emergence and
maintenance of meaning through successful interaction and its
knowledge base within the brain. By repeated trial-and-error each
brain constructs within itself an understanding of its surrounding,
which constitutes its {\it knowledge} of its own world that we
describe as its {\it Double} \cite{Vitiello:2001}. It is an {\it
active} mirror, because the environment impacts onto the self
independently as well as reactively. The notion of an order
parameter denotes a categorial descriptor that exists only in the
brain, so that its matching "double" is a finite projection from the
brain into the environment, as the basis for organizing the action
of the body governed by the brain. An example is the grasping of an
object by the hand,  described by the phenomenologist Merleau-Ponty
\cite{MerleauP:1945a} as the achievement of "maximum grip". Thus we
conceive the "double" as the descriptor of the perception or
experience of the object, as contrasted with the brain activity
pattern that is matched by the "double". Such a matching is formally
described by the continual balancing of the energy fluxes at the
brain--environment interface. It amounts to the continual updating
of the {\it meanings} of the flow of information exchanged in the
brain behavioral relation with the environment.

Perhaps, at the present status of our research, we might conclude
that the dissipative quantum dynamics underlying textured AM
patterns and sequential phase transitions observed in brain
functioning could open the way to  understand John von Neumann's
remark: ``...the mathematical or logical language truly used by the
central nervous system is characterized by less logical and
arithmetical depth than what we are normally used to. ...We require
exquisite numerical precision over many logical steps to achieve
what brains accomplish in very few short steps" (pp.80-81 of
\cite{vonNeumann:1958a}).

\ack The authors thank Prof. Mariano A. del Olmo and the organizers
of the International Conference ``Quantum Theory and Symmetries,
QTS-5", held in Valladolid, Spain, July 2007,  for giving them the
opportunity to present in that Conference the results reported in
this paper. Partial financial support from MUR and INFN is also
acknowledged.

\appendix

\section{Spontaneous breakdown of symmetry in quantum field theory}

Symmetry is said to be spontaneously broken when the Lagrangian of a
system is invariant under a certain group of continuous symmetry,
say $G$, and the vacuum or ground state of the system is not
invariant under $G$, but under one of its subgroups, say $G'$
\cite{Umezawa:1993yq,ITZ,Anderson:1984a}. The ground state then
exhibits observable ordered patterns corresponding to the breakdown
of $G$ into $G'$ \cite{Umezawa:1993yq,Anderson:1984a,Marshak}. The
possibility of having different vacua with different symmetry
properties is provided by the mathematical structure of QFT, where
infinitely many representations of the canonical commutation
relations (CCR) exist, which are unitarily inequivalent  with
respect to each other, i.e there is no unitary operator transforming
one representation into another one \cite{68}, and thus they are
physically inequivalent as well: they describe different physical
phases of the system. By contrast, in Quantum Mechanics all
representations are unitarily (and therefore physically) equivalent
\cite{Umezawa:1993yq,vonNeumann,QM}.

In SBS theories, the Goldstone theorem predicts the existence of
massless bosons, called  Nambu-Goldstone (NG) particles
\cite{Goldstone}. The spin-wave quanta, called magnons in
ferromagnets, the elastic wave quanta, called phonons in crystals,
the Cooper pair quanta in superconductors, etc.
\cite{Umezawa:1993yq,Anderson:1984a}, are examples of NG particles.
NG bosons condensed in the ground state of the system according to
the Bose-Einstein condensation  are the carriers of ordering
information out of which ordered patterns (space ordering or time
ordering as, e.g., ``in phase" oscillations) are generated. The
condensation density of the NG boson quanta determines the
macroscopic field which is called {\it order parameter}, e.g. the
magnetization in ferromagnets. The order parameter is a classical
macroscopic field in the sense that it is not affected by quantum
fluctuations. Its value may be considered to be the {\it code} or
{\it label} specifying the physical phase of the system.

In the absence of gauge fields, the NG quanta are observed as
realistic physical quanta, and  excitations of the vacuum extend
over the whole system ({\it collective modes} or {\it long range
correlations}). They may scatter with other particles of the system
or with observational probes. If a gauge field is present, the NG
bosons still control the condensation in the ordered domain, and the
gauge field propagation is confined into regions where the order is
absent (e.g. in the core of the vortex in superconductors,
Anderson-Higgs-Kibble mechanism)
\cite{Umezawa:1993yq,Anderson:1984a,Higgs,Leggett:1980}.

Through the generation of NG collective modes, SBS is responsible
for the change from microscopic to macroscopic scale
\cite{Umezawa:1993yq,Anderson:1984a}: crystals, ferromagnets,
superconductors, etc. are {\it macroscopic quantum systems}. They
are quantum systems not in the sense that they are constitued by
quantum components (like any physical system), but in the sense that
their macroscopic properties, accounted for by the order parameter
field, cannot be explained without recourse to the underlying
quantum dynamics.

We finally comment on the Hermitian conjugation of the Hamiltonian
in the real time finite temperature formalism (thermo field dynamics
(TFD)), where there are three free parameters, $f,~\alpha,~s$, in
the notation of \cite{Umezawa:1993yq}, corresponding to the three
parameters of the $SU(1,1)$ group.  The parameter $s$ does not
contribute to the propagator \cite{Umezawa:1993yq} and is usually
set equal to zero since no physical meaning is attached to it.
$\alpha$ is related to the cyclic  property of the trace operation
$Tr[\rho A] = Tr[\rho^{1-\alpha} A \rho^{\alpha}]$. Physical
observables are independent of $\alpha$. The choice $\alpha = 1$ (or
$\alpha = 0$) turns out to be convenient in Feynman graph
computations in non-equilibrium TFD \cite{TakUmez,Umezawa:1993yq}.
The choice $\alpha = \frac{1}{2}$ preserves the usual definition of
Hermitian conjugation. Other choices give  so-called non-Hermitian
representations of TFD. Since the physical content of the model is
not affected, we use $\alpha = \frac{1}{2}$, as far as we are not
involved in computations of Feynman graphs. The parameter $f$ is the
only physically relevant parameter. It is related to the canonical
Bose distribution (\ref{bose}), in which case it is $f = e^{- \beta
E}$, and thus it determines the ${\cal N}_{A_{\kappa}}$s.

\section{Neurophysiological observations}

The tight sequencing of AM patterns requires high temporal
resolution.  Hilbert transform is then applied to EEG signals after
band pass filtering  \cite{7,9,10}. Unlike the Fourier transform
that decomposes an extended time series into fixed frequency
components, the Hilbert transform decomposes an EEG signal into the
analytic amplitude $A(t)$,  the analytic phase $P (t)$ and the
instantaneous frequency, $\omega (t)$, at each digitizing time step
on each channel.

The analytic phase difference $\Delta P_{j}(t) = P_{j} (t)-P_{j}
(t-1)$ at each electrode and at each digitizing step divided by  the
digitizing time increment   specifies the instantaneous frequency:
$\omega_{j} (t) = \frac{\Delta P_{j}(t)}{\Delta t}$. It has been
shown in \cite{9} that the rate of increase in phase (the mean
instantaneous frequency $= 0.4 ~rad/2 ms = 31 ~Hz$) is relatively
constant in epochs that last $ \sim 60-100 ~ms$ and that recur at
intervals in the theta and alpha ranges.  These plateaus in nearly
constant phase increase are bracketed by phase discontinuities
synchronized across the array \cite{9}. This spatially correlated
'phase slip' demarcates {\it phase transitions} in the cortical
dynamics. The brackets are detected and displayed as spikes (see
Fig. B1) by calculating the spatial standard deviation of the phase
differences, $SD_{X}(t)$, across the array as a time series for the
$64$ signals. $SD_{X}(t)$ is thus a useful index of the temporal
stability. The spikes bracket the stabilized epochs and define the
beginning and end of wave packets; the plateaus demarcate epochs of
near stationarity.

Calculation of $SD_{X}(t)$, and the mean analytic amplitude
$\underline{A}(t)$ across $n$ channels at each time point confirmed
\cite{9, 10, 27} that peaks in $\underline{A}(t)$ accompany plateaus
in $SD_{X}(t)$ (Fig. B1). Peak amplitudes enable optimal measurement
of spatial patterns of AM of beta or gamma carrier frequency. Each
pattern is expressed by an $n \times 1$ feature vector in the square
of amplitude, $A^{2}_{j}(t)$. The mean power,
$\underline{A}^{2}(t)$, serves as a scalar label for each AM
pattern.

These broad AM patterns are the neural correlates to display the
rapid re-organization of brain activity that we believe underlies
both cognitive function and sequences of complex intentional
behaviors. Owing to the potential differences that dendritic
currents maintain as they flow across the relatively fixed
extracellular impedance of the neuropil, the values of
$A^{2}_{j}(t)$  provide a measure of the rates of free energy
dissipation required by the neurons generating the ECoG. Our index
of those energy levels may be optimally correlated with patterns of
increased blood flow that indirectly manifest the metabolic energy
utilization by parts of the brains, which are detected with fMRI,
PET,  and SPECT \cite{41}.

At first view the AM patterns appear to be ``cortical
representations" of conditioned stimuli (CS). However, the patterns
that are elicited by an invariant CS hold only within each training
session and then only if there are no changes in the schedule of
reinforcement or addition of a new CS in serial conditioning.
Measurements of AM patterns within sessions show pattern variation
within each category despite CS invariance. Between sessions with no
new CS added the averages of the patterns tend to drift. When the
subjects are trained to respond to a new CS, all of the patterns
change, including the pattern for the background. The amount of
change with new learning is $2$ to $4$ times the average change with
drift across multiple sessions \cite{80,33,35}. A collection of AM
patterns that we established by training persisted with drift
through multiple sessions until we introduced the next contextual
change.

Every AM pattern is accompanied by a conic phase pattern that
retains the history of its site of nucleation and spread. Phase
cones were also found between ordered frames and overlapping with
them at near and far frequencies. In a distributed medium such as
the neuropil, the generation of the cortical standing wave resulting
from  a phase transition forming a wave packet begins at a site of
nucleation and spreads radially at a velocity determined by the
propagation velocities of axons extending parallel to the surface.
This gives a conic phase gradient and the illusion of a traveling
wave by the delay in initialization embodied in the phase cone. This
is measured by fitting a cone to a phase surface given by the
analytic phase, $P_{j}(t), ~ j = 1,...,64$. The phase transitions
appear to be induced by input to the cortex serving as a control
parameter; however, the latency varies randomly with respect to
known times of input onset.

\begin{figure}
\begin{center}
\includegraphics [width=20pc] {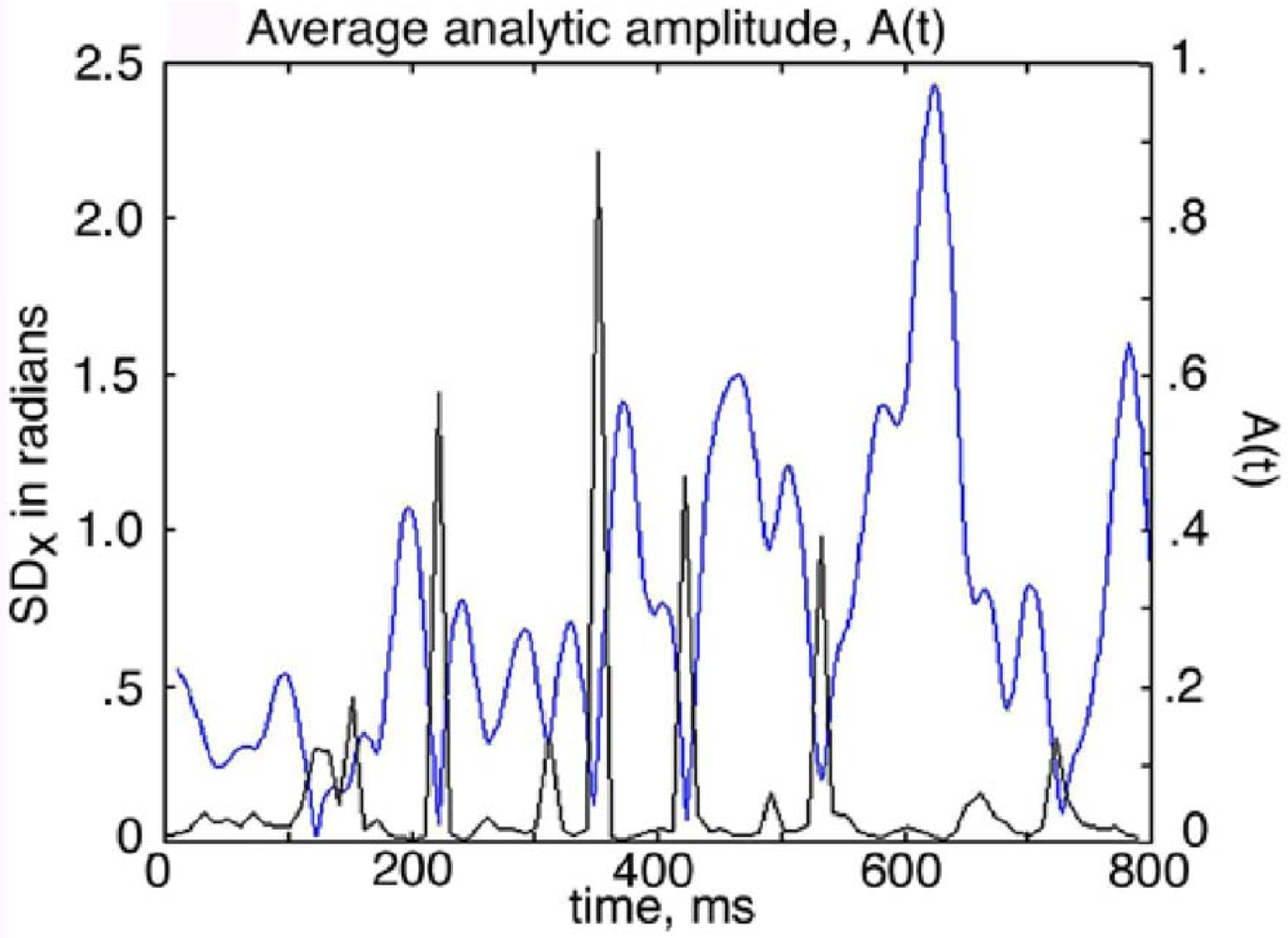}
\end{center}
\caption{\nonumber \small \noindent  
The analytic amplitude, A(t), of the ECoG in the beta band
fluctuates with time. The maxima are textured with AM spatial
patterns. The minima are accompanied by spikes in the spatial
standard deviation of the phase differences as a function of time,
SDx(t). Each spike reflects the indeterminacy of phase at the null
spike in amplitude, where a phase transition is enabled.}
\end{figure}

On successive trials with the same CS the location of the apex
varies randomly within the primary receiving area for the CS
modality, and its sign (maximal lead as in an explosion or maximal
lag as in an implosion) likewise varies randomly from each phase
transition to the next. These random variations give further
evidence for SBS \cite{11X}.

\section*{References}


\end{document}